\documentclass{aastex63}

\received{June 1, 2019}
\revised{January 10, 2019}
\accepted{\today}
\submitjournal{APJ}

\graphicspath{{./}{figures/Results Plots/}}
\usepackage{url}
\usepackage{amssymb}
\usepackage{dsfont}
\usepackage{amsmath}
\usepackage{float}
\usepackage{longtable}

\begin{document}

\title{Predicting the redshift of gamma-ray loud AGNs using Supervised Machine Learning: Part 2}

\author{Aditya Narendra}
\affiliation{Jagiellonian University, Poland}

\author{Spencer James Gibson}
\affiliation{Carnegie Mellon University, USA}

\correspondingauthor{Maria Giovanna Dainotti}
\email{maria.dainotti@nao.ac.jp}

\author{Maria Giovanna Dainotti}
\affiliation{National Astronomical Observatory of Japan, Mitaka}
\affiliation{Space Science Institute, 4750 Walnut St, Suite 205, Boulder,CO,80301,USA}

\author{Malgorzata Bogdan}
\affiliation{Department of Mathematics, University of Wroclaw, Poland}
\affiliation{Department of Statistics, Lund University, Sweden}

\author{Agnieszka Pollo}
\affiliation{Astronomical Observatory of Jagiellonian University, Krakow}
\affiliation{National Centre for Nuclear Research, Warsaw}

\author{Ioannis Liodakis}
\affiliation{Finnish Centre for Astronomy with ESO (FINCA), University of Turku, Finland}

\author{Artem Poliszczuk}
\affiliation{National Centre for Nuclear Research, Warsaw}





\begin{abstract}

\textcolor{black}{
Measuring the redshift of active galactic nuclei (AGNs) requires the use of time-consuming and expensive spectroscopic analysis.}
{However, obtaining} \textcolor{black}{redshift} measurements of AGNs is \textcolor{black}{crucial as it can enable AGN population studies, \textcolor{black}{provide} insight into the star formation rate, the luminosity function\textcolor{black}{, and} the density rate evolution}. 
\textcolor{black}{Hence, there is a requirement for alternative redshift measurement techniques}.
\textcolor{black}{
In this project, we aim to use the \textcolor{black} {{\it Fermi} gamma-ray space telescope's} 4LAC Data Release 2 (DR2) catalog to train a machine learning model capable of predicting the redshift reliably.}
\textcolor{black} {In addition, this project aims at improving and extending} with the new 4LAC Catalog \textcolor{black}{the predictive capabilities of the machine learning (ML) methodology published in \cite{dainotti2021predicting}. 
Furthermore, we implement feature engineering to expand the parameter space and a bias correction technique to our final results. 
\textcolor{black}{
This study, uses additional machine learning techniques inside the ensemble method, the SuperLearner, previously used in \cite{dainotti2021predicting}.}
Additionally, we also test a novel ML model called Sorted L-One Penalized Estimation (SLOPE).
Using these methods we provide a catalog of estimated redshift values for those AGNs that do not have a spectroscopic redshift measurement.
These estimates can serve as a \textcolor{black}{redshift} reference for the community to verify as updated {\it Fermi} catalogs are released with more redshift measurements.}

{
}
\end{abstract}



\section{\textbf{Introduction}} \label{sec:intro}


\textcolor{black}
{
The {\it Fermi} $\gamma$-ray space telescope  observes the sky in the energy range of 50 MeV to 1 TeV. With its $\gamma$-ray observational capabilities, it has observed 3511 $\gamma$-ray loud active galactic nuclei (AGNs) to this day.
These AGNs are gathered in the latest Fourth {\it Fermi}-LAT Data Release 2 (4LAC) catalog \citep{2020ApJ...892..105A,2020ApJS..247...33A}, along with their measured $\gamma$-ray properties.
The majority ($>98\%$) of AGNs in the 4LAC catalog are blazars, which according to the Unified Model of AGN, are identified with galaxies that have their relativistic emission jets pointed directly along our line of sight~\citep{Blandford2019}.
Two major subcategories of blazars are present in the 4LAC catalog, BL Lacertae Objects (BLL) and Flat Spectrum Radio Quasars (FSRQ). 
BLLs are classified as sources with weak or no emission lines in their spectra, whereas FSRQs are classified as sources that have broad emission lines ~\citep{Madejski2016ARA&A_gammaAGN,ghisellini2011transition}.
}





{Measuring the redshift of blazars proves to be problematic as it often requires multiwavelength observations and coordination across multiple facilities. \textcolor{black}{Further complications arise because the most dominant category of $\gamma$-ray loud blazars, BLLs, often do not show detectable emission lines.}
\textcolor{black}{Not surprisingly, only 50\% of $\gamma$-ray AGNs have redshift estimates}, highlighting the difficulty in measuring it, as well as the need for a reliable photometric redshift estimation method.
In light of these hurdles, a predictive approach for redshift estimation using the more readily available photometric $\gamma$-ray features is an attractive proposition.
}

\begin{figure}[H]
    \centering
    \includegraphics[width=0.8\textwidth]{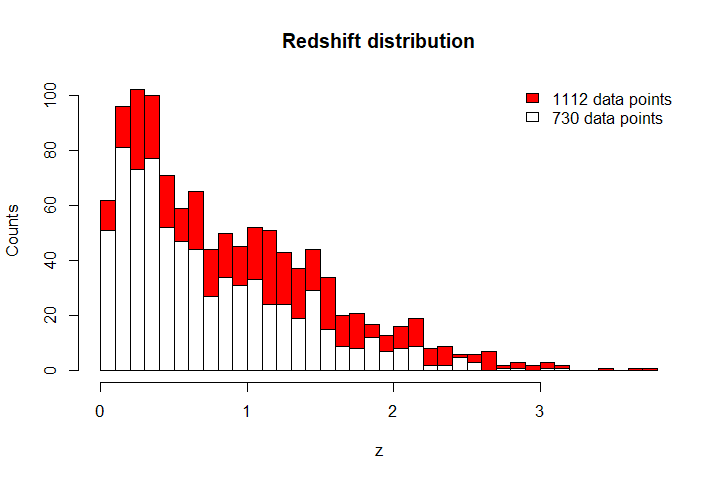}
	\caption{The histogram showing the overlapped distributions of z in \cite{dainotti2021predicting} training data set (white) and the current training data set (red). }
    \label{fig:red2}
\end{figure}

\textcolor{black}{
Photometric redshifts \textcolor{black}{can} play a significant role in helping us understand the origin of Extragalactic Background Light (EBL),
\textcolor{black}{crucial for blazars emitting GeV-TeV $\gamma$-rays, which are typically BLLs \citep{Wakely2008}\footnote{http://tevcat.uchicago.edu/}.}
This can, in turn, enhance our knowledge about the cosmic evolution of blazars \citep{singal2012flux,singal2014gamma,singal2015determination,singal2013flat,chiang1995evolution,ackermann2015multiwavelength,singal2013cosmological,ackermann2012gev}, the structure of the magnetic field in the intergalactic medium, \citep[][]{2020AAS...23540506M,2013MNRAS.432.3485V,2018Sci...362.1031F}
as well as constraining cosmological parameters \citep{2019ApJ...885..137D,petrosian1976surface,singal2013cosmological}.
}
\textcolor{black}{
In recent years the number of studies focusing on photometric redshift estimation of high redshift AGN, using a machine learning (MLapproach, has increased significantly \citep{jones2017analysis,cavuoti2014photometric,fotopoulou2018cpz,logan2020unsupervised,yang2017quasar,zhang2019machine,curran2020qso,  nakoneczny2020photometric,pasquet2018deep,jones2017analysis}. 
This is primarily due to the availability of large data sets from all-sky surveys like the Sloan Digital Sky Survey (SDSS) \citep{aihara2011eighth} and Wide-field Infrared Survey Explorer (WISE) \citep{Brescia2019,ilbert2008cosmos,hildebrandt2010phat,brescia2013photometric,2010,DIsanto2018}. 
However, when it comes to $\gamma$-ray loud AGNs, ML methods have been primarily used for AGN classification \citep{doert2014search,hassan2013gamma,einecke2016search,lefaucheur2017research,kang2019evaluating,Chiaro2016,Liodakis2019}.
\newline\indent
\cite{dainotti2021predicting} remains the only work in the field, as of this writing, that investigates the capabilities of {\it Fermi} in estimating photometric redshifts.
However, in that study, a significant hindrance was the small dataset, which also contained a large number of missing values. 
Since AGNs need to have both observed spectroscopic redshift and complete observational data to be included in the ML models, this led to only 50\% of the data set being eligible. As a result, the small training sample contained 730 AGNs.
Hence, in this research one of our primary focuses was on increasing the training sample size, which was achieved by including AGNs from the 4LAC catalog that are present below the galactic plane.
A detailed description of the differences between the data set of the current work and \cite{dainotti2021predicting} is provided in Sec. \ref{sec:conclusion}.
\newline\indent
We also increase the number of predictors using feature engineering techniques (see Sec. \ref{sec:featureEngineering}). 
We also use additional ML models in our ensemble (see Sec. \ref{sec:ML}), chosen specifically for this updated data set, and an independent method called Sorted L-One Penalized Estimation (SLOPE) (see Sec. \ref{sec:SLOPE}).
Furthermore, we investigate the effects of 
bias correction methods (see Sec. \ref{sec:BiasCorrection}) on our results. 
Finally, we present a list of photometric redshift predictions for the AGNs in the 4LAC that do not have a spectroscopic redshift measurement (see Sec. \ref{sec:appendix}). 
\newline\indent
With these updates, we were able to maintain low numbers of catastrophic outlier percentage (5\%), small values of the root mean square error (RMSE), and the normalized median absolute deviation ($\sigma_{NMAD}$) as in \cite{dainotti2021predicting}. 
We also managed to increase the Pearson correlation value, r, between our predicted and observed redshifts.
The results on the validation set
are also improved by 10\%.
We here stress that this has been achieved on training and validation data samples 52\% larger than \cite{dainotti2021predicting} (see Fig. \ref{fig:red2}). 
}

\section{\textbf{The sample}} \label{sec:sample}
\textcolor{black}{
Since 2008} \textcolor{black}{ the instruments onboard {\it Fermi} (LAT, GBM) have been continuously detecting $\gamma$-ray sources (e.g., Gamma-Ray Bursts, AGNs, among others).}
\textcolor{black}{Out of the 3511 AGNs observed in the latest 4LAC catalog, only 1767 have measured redshifts, 50.3\% of the total data set. 
In the 4LAC catalog, two categories of $\gamma$-ray loud AGNs dominate, BLLs and FSRQs. The catalog contains a total of 1190 BLLs and 733 FSRQs. 
A significant percentage (43\%) of the AGNs are unclassified and are denoted as Blazar Candidate of Unknown Type (BCU). 
We also have 9 Narrow Line Seyferts (NLSY1), 44 radio galaxies, and 19 AGNs of other categories.
Keeping this analysis consistent with the work previously done in \cite{dainotti2021predicting}, and to improve the readability and interpretability of the method, we choose to omit all the non-BLL and non-FSRQ type AGNs from our data.
}

\begin{figure}
    \centering
    \includegraphics[width=0.8\textwidth]{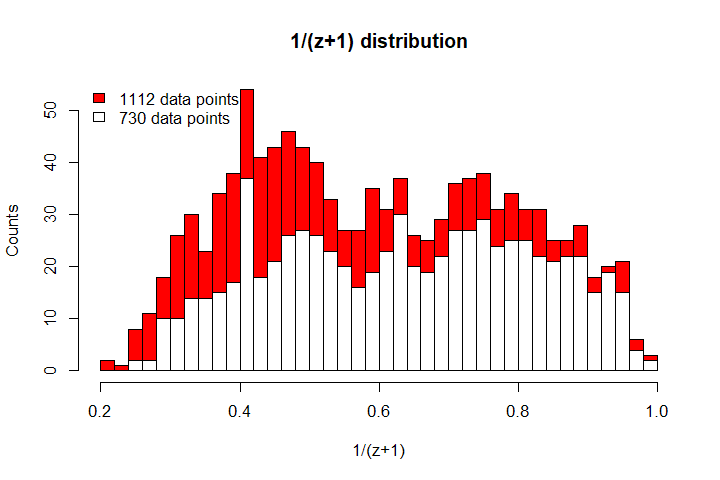}
	\caption{The overlapped histogram between \cite{dainotti2021predicting} data set and the current one, in the $1/(z+1)$ scale. }
    \label{fig:invred}
\end{figure}

\textcolor{black}{
Furthermore, Fermi measures 13 properties of the 4LAC AGNs. These observations are used as predictors for the redshift in our ML models.
The corresponding Gaia g-band magnitudes \citep{2010A&A...523A..48J} of the AGNs are also included. 
Out of these 13 features or predictors, we consider 11,  excluding $Fractional\_Variability$ and $Highest\_Energy$. 
Both of them were excluded due to a lack of observational data in 43\% of the AGNs.  
The definitions of the 11 predictors are provided below.
}


\begin{itemize}\label{best}
\item \textit{Flux} - The integral photon flux, in photons/cm2/s, from 1 to 100 GeV.
\item \textit{Energy\_Flux} - The energy flux, the units are in erg cm$^{-2}$ s$^{-1}$, in the 100 MeV - 100 GeV range obtained by the spectral fitting in this range. 
\item \textit{Significance} - The source detection significance in Gaussian sigma units, on the range from 50 MeV to 1 TeV. 
\item \textit{Variability\_Index} - Summation of the log(likelihood) differences between the fitted flux in every interval of time and the mean flux over the 50 MeV to 1 TeV range. 
\item \textit{$\nu$} - The synchrotron peak frequency, measured in Hz, in the observer frame.
\item \textit{$\nu$f$\nu$} - Spectral energy distribution at the synchrotron peak frequency in the $\nu$f$_\nu$ spectrum.
\item \textit{PL\_Index} - It is the photon index when fitting the spectrum with a power law, in the energy range from 50 MeV to 1 TeV.
\item \textit{Pivot\_Energy} - The energy, in MeV, at which the error in the differential photon flux is minimal, is derived from the likelihood analysis in the range from 100 MeV - 1 TeV.
\item \textit{LP\_Index} - Photon index at pivot energy ($\alpha$) when fitting the spectrum (100 MeV to 1 TeV) with Log Parabola. 
\item \textit{LP\_$\beta$} - the spectral parameter ($\beta$) when fitting with Log Parabola spectrum from 50 MeV to 1 TeV.
\item \textit{Gaia\_G\_Magnitude} - Gaia Magnitude at the g-band provided by the 4LAC, taken from the Gaia Survey.
\end{itemize}


\begin{figure}[H]
    \centering
	\includegraphics[width=0.99\textwidth]{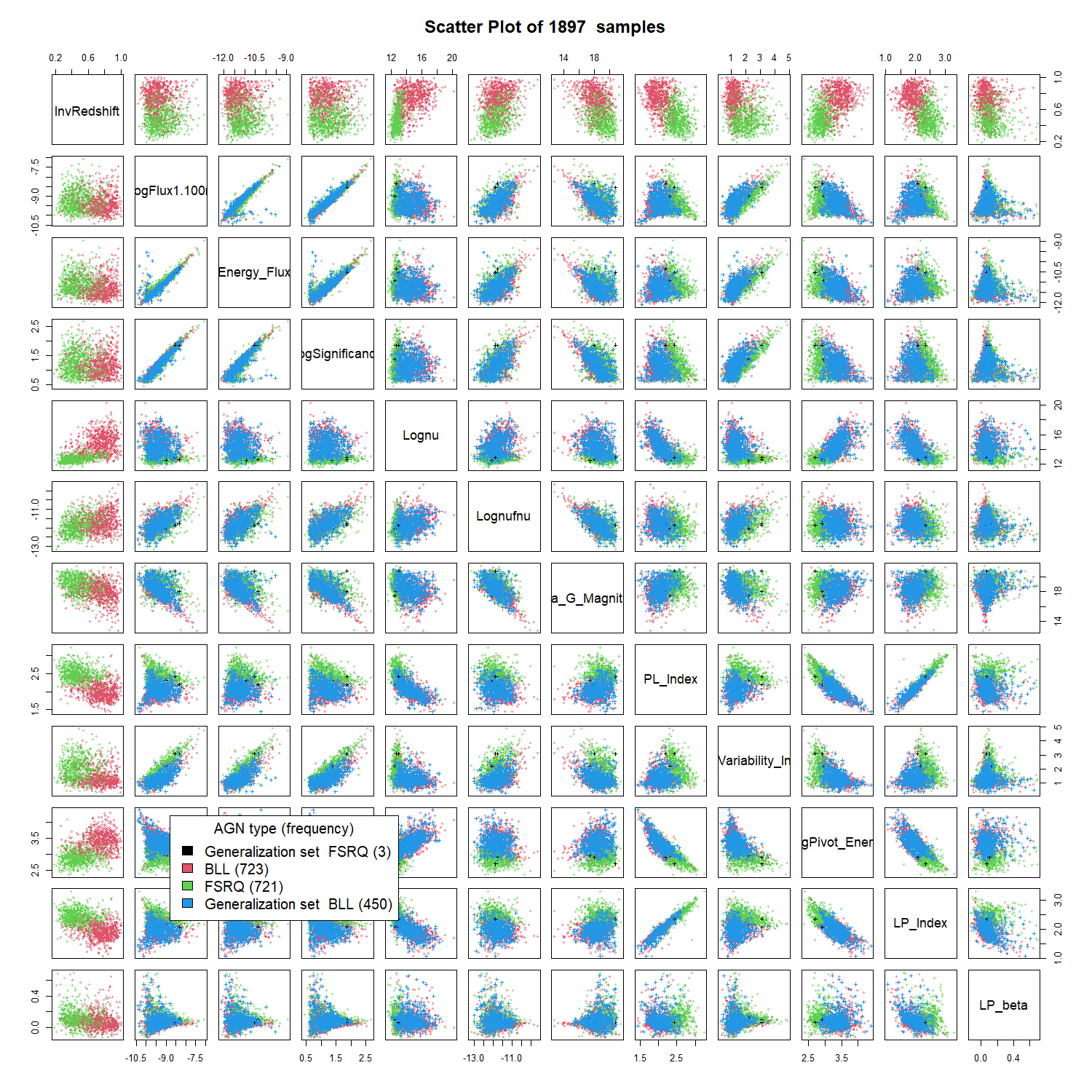}

    \caption{The full scatter matrix plot of all the variables defined above. Here \textit{InvRedshift} denotes $1/(z+1)$ scaled data. The generalization set refers to samples which do not have a redshift observation.}
    \label{fig:fullscatter}
\end{figure}

\subsection{Training, generalization and validation sets}\label{sec:sets}

\textcolor{black}{
Before we can train the ML models, we need to transform our predictors and create the training and generalization sets.}
For the predictors' $Flux, Energy\_Flux, Significance, Variability\_Index, \nu, \nu$f$\nu $, and $ Pivot\_Energy$, a transformation to their logarithmic scale is necessary as they span over several orders of magnitude.
Also, note that we do not train our ML models to predict the redshift directly, instead, it is trained to predict $1/(z+1)$, where z is the redshift.
\textcolor{black}{With such a transformation, we predict the scale factor $a$, which is equal to $1/(z+1)$; furthermore, we found this transformation to be better for the overall performance of our ML models.}
The 1/(z+1) is indeed a much more physical quantity to estimate, as it is directly used for cosmological and redshift evolution studies in both AGNs and GRBs. \citep{dainotti2013determination,dainotti2017study,dainotti2021hubble,singal2015determination,2013MNRAS.436...82D,2013ApJ...774..157D,2021ApJ...912..150D,2017NewAR..77...23D,2018PASP..130e1001D}. 
\textcolor{black}{
\newline\indent
Furthermore, we also make cuts in the sample to remove data points with an $LP\_\beta$ $<$ 0.7, $LP\_Index$ $>$ 1, and $LogFlux$ $>$ -10.5, as they are outliers of the distribution.
These cuts leave us with 1897 AGNs, out of which 1444 have observed redshift and the remaining 453 do not. 
The scatter matrix plot in Fig. \ref{fig:fullscatter} shows the distribution of these 1897 samples.
}

\textcolor{black}{
The training set or the training sample is the data on which our ML models are trained to learn the underlying relations between the response variable and the predictors.
Thus, it consists of AGNs with observed redshift (the response variable) and predictors with complete data.}
{Due to missing values in 3 predictors, namely $Log\nu$, $Log\nu f\nu$, and $Gaia\_G\_Magnitude$, we had to remove an additional 332 AGNs.
This leads to a training sample of 1112 data points, which is 52\% larger than \cite{dainotti2021predicting}. 
The overlapped histogram of the redshifts of the entire data and the training data is presented in Fig. \ref{fig:redshiftOverlap}.
\begin{figure}
    \centering
	\includegraphics[width=0.8\textwidth]{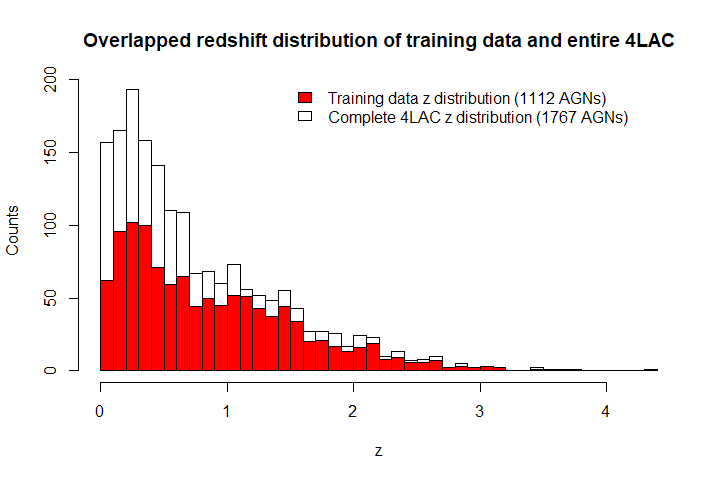}
    \caption{Overlapped histogram showing the training set and the total 4LAC redshift distributions.}
    \label{fig:redshiftOverlap}
\end{figure}
\newline\indent
On the other hand, the generalization set contains AGNs without observed redshift but with complete observational data. 
These are the AGNs for which we want to estimate the photometric redshift. 
Similar to the training set, if we remove the data points with missing values then we are left with 320 AGNs in the generalization sample. These 320 AGNs are made up of 319 BLLs and 1 FSRQ.
\newline\indent
The validation set consists of AGNs, which contain measured redshift and complete predictor data similar to the training set. 
Hence, we take a cut of 10\% from the training set to create the validation set.
This set of 111 AGNs will not be used in the model's training and will serve as an independent test to verify the cross-validated results.
}

\section{\textbf{Methodology}} \label{sec:methods}

\textcolor{black}{
For this project, we expand upon the methodology published in \cite{dainotti2021predicting}, which includes a more accurate version of feature selection where we select the best predictors for estimating $1/(z+1)$, additional ML models, feature engineering, 
and bias corrections.
All the techniques and methods are described in the following subsections. 
}


\textcolor{black}{The metrics we use for measuring the goodness of our results are the following:}
\begin{itemize}
    \item Bias: Mean of the difference between the observed and predicted values.
    \item $\sigma_{NMAD}$: Normalized median absolute deviation between the predicted and observed measurements.
    \item Pearson correlation coefficient, $r$, between the predicted and observed measurements.
    \item Root Mean Square Error (RMSE).
    \item Standard Deviation $\sigma$.
\end{itemize}
We quote the measured values of these parameters for $\Delta z_{norm}$ and $\Delta z$. Where:
$$
\Delta z = z_{observed} - z_{predicted} 
$$
$$
\Delta z_{norm} = \frac{\Delta z}{1 + z_{observed}}
$$
Along with these we also quote the catastrophic outlier percentage, which is defined as the percentage of predictions that lies beyond the 2$\sigma$ error.
\textcolor{black}{The quoted metrics used here are the same as in \cite{dainotti2021predicting} and other works in the field, thus allowing for easy comparison.}

\subsection{Cross-Validation}\label{sec:CV}

{
{Cross-validation (CV) is a frequently used method \citep{browne2000cross,refaeilzadeh2009cross} which estimates the expected prediction error of ML models, which in our case is the RMSE.
Commonly, a $10$-fold cross-validation (10fCV) is used.
In 10fCV, ten equal partitions or folds are created at random.
At each step, nine folds are used as training data to fit a model and the 10th fold is used as the test set, on which the prediction error is evaluated. 
This process is iterated such that 
each fold is used exactly once as a test set.
The averaged errors of each test set provide a metric to judge the performance of the ML model on novel data samples.
}
\textcolor{black}{Since the folds are generated randomly, there is a risk of over-fitting to a specific data partition. Therefore, we perform 10fCV 100 times, and for each iteration, the folds are generated randomly.
The results are an average of these 100 iterations. Such a step is necessary to mitigate the uncertainty of 10fCV and derandomize our results.  
However, it should also be noted that this step is computationally expensive, as we need to train multiple ML algorithms 10$\times$100 = 1000 times. 
}

}

\subsection{Feature selection : LASSO inside 10fCV}\label{sec:FeatureSelection}
\textcolor{black}{One of the most crucial steps in any ML project is feature selection. It is the process where we choose a subset of all the predictors present in the data set that are better at predicting the response variable.
Feature selection is crucial because it makes the model easier to interpret and reduces the size of the high-dimensional data, which enables algorithms to work faster.
As a consequence, this also reduces the risk of over-fitting.
Here we use the Least Absolute Shrinkage and Selection Operator (LASSO) method \citep{TibshiraniLasso}. 
\newline\indent
The LASSO algorithm works as a shrinkage method by constraining the $\ell^1$-norm of the solution vector to be less than or equal to a positive number known as the tuning parameter ($\lambda$). 
A vector's $\ell^1$-norm is defined as the sum of the magnitudes of the components of that vector.
Such a constraint on the solution vector penalizes the predictor coefficients, resulting in some of the coefficients being set to 0, effectively eliminating those predictors. 
The $\lambda$ parameter controls the shrinkage that is applied to the coefficients of the predictors. 
}
\textcolor{black}{Our analysis, uses the R implementation of the GLMNET function and the LASSO feature selection mode by setting the parameter $alpha$ to 1 \citep{hastie2017extended,tibshirani2012strong}.
The function varies $\lambda$ while fitting a linear model to the data and assigns coefficients to the predictors to reduce the error.
As mentioned previously, the actual features that are chosen depending on the value of $\lambda$. 
Thus, we select those predictors with a non-zero coefficient for the highest $\lambda$ value such that the prediction errors are within 1 standard deviation  \citep{hastieTibs,birnbaum1962foundations,hastie1987generalized,hastie1990generalized,friedman2010regularization}.
}

{In this project, 
we perform LASSO feature selection 
during the cross-validation step. 
This means that instead of choosing a feature set for the entire training data, we pick optimal features from the data of the nine folds that will be used to train the model at that stage. 
Then the model predicts on the test set (10th fold) using these features only, such that each of the ten folds acts as a test set exactly once.
Note that the LASSO feature selection technique uses the response variable to choose the optimal subset of predictors.
Hence, this method assures us that we are not using information from the test set of the 10fCV, and allows us to substantiate the claim that the CV results accurately estimate the expected prediction error for this method. 
}

\subsection{The ML algorithms used in our analysis}\label{sec:ML}

\begin{figure}
    \includegraphics[width = 0.49\textwidth]{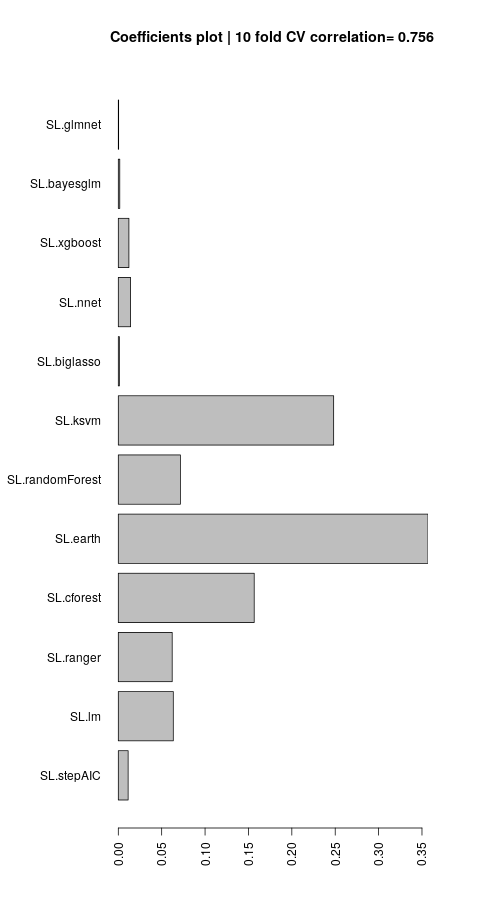}
    \includegraphics[width = 0.49\textwidth]{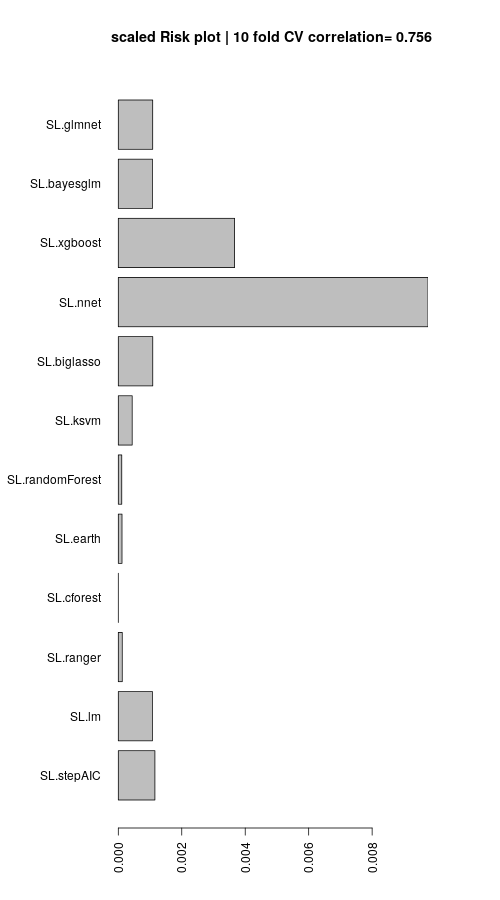}
    \caption{
    Left Panel: The coefficients assigned by SuperLearner to the algorithms tested. We select the algorithms that have a coefficient above 0.5 to be incorporated into our ensemble. 
    Right Panel: The RMSE error (risk) of each of the algorithms, scaled to show the minimum risk algorithm at 0, which is cforest.}
    \label{fig:all_algo}
\end{figure}

\textcolor{black}{
The choice of the best ML algorithms depends significantly on the problem being addressed and the data set being used. That holds in our study as well.
Previously in \cite{dainotti2021predicting}, we tested multiple ML algorithms in an ensemble and selected the ones with the highest coefficients. 
We perform the same process once again, but with an expanded data set and different predictors. 
Unsurprisingly, different ML algorithms were chosen to be best suited for this analysis compared to \cite{dainotti2021predicting}. 
In Fig. \ref{fig:all_algo}, we present the coefficients assigned to the various ML models we tested. 
Those with a higher coefficient are better at predicting the redshift. Therefore, we define a threshold value of 0.5 and choose all the models with a coefficient greater than this. 
Consequently, we end up with six algorithms in our final ensemble. These are Enhanced Adaptive Regression Through Hinge (EARTH), KSVM, Cforest, Ranger, Random Forest, and Linear Model. 
We describe these methods below.
}

\subsubsection{EARTH}
{Enhanced Adaptive Regression Through Hinges (EARTH) is an implementation of the multivariate adaptive regression splines method (MARS), based on the work by \cite{FriedmanMARS}. 
MARS is a non-parametric regression technique that can better model the interactions of predictors and non-linearities than a simple linear model.
MARS starts with a constant term then fits a weighted sum of functions where each function is either another constant term, a hinge function, or a product of hinge functions. 
Each term added to the model results in the maximal reduction of sum-of-square residual error.
}

\subsubsection{KSVM}
{In this study, we use the support vector regression method in the ksvm algorithm from the SuperLearner package. 
Support vector regression (SVR) is the continuous predictive counterpart to support vector machine classification~\citep{cortes1995support}. 
SVR uses a kernel function that maps an input vector onto a space of higher-dimension.
Then it fits a hyperplane to the data, such that the predictions are within a certain error margin.
In this study, the default parameters of the ksvm model in the SuperLearner package are used. 
Specifically, the kernel used is the radial basis kernel, i.e. the Gaussian kernel.
} 

\subsubsection{Cforest, Ranger, and Random Forest}
{The Cforest, Ranger, and Random Forest algorithms belong to the tree class of supervised machine learning models. 
\textcolor{black}{A single decision tree algorithm may perform very well on the training data, but it is prone to overfitting. Besides the true relationships, such models learn the noise present in the training set and are referred to as models with high variance. 
To obtain an optimal bias-variance trade-off, one can often use ensembles made of multiple decision trees. 
In such models, several stages of randomness are introduced to reduce the variance of the final prediction.
}
\newline\indent
Random Forest~\citep{breiman2001random} works by generating multiple \textcolor{black}{decision} trees. 
While building the tree the data is partitioned based on the values of a randomly chosen subset of features in the training data \citep{randomForestsHo}.
This partitioning repeats until the pre-specified depth of the tree is reached. 
At the nodes, the values of the response variables are averaged over and that average is assigned as the prediction for an AGN that falls into that node.
The algorithm finally outputs the prediction as an average over all the independent trees.
} \par

{The Ranger algorithm is nearly equivalent to Random Forest with the difference being a slightly quicker implementation and the ability to use extremely randomized trees (ERT). 
ERTs are built in the same top-down procedure as decision trees in Random Forest with the difference being that the cuts are randomly generated, assuming a uniform distribution for each of the features in the training set.
The cut that reduces prediction error the most is chosen \citep{geurts06extremetrees}. 
ERTs have the advantage of reducing variance further than Random Forest due to the increased randomness in the node-splitting procedure. 
\newline\indent
The Cforest algorithm \citep{hothorn2006unbiased} is similar to Random Forest and Ranger except that it uses conditional inference trees \citep{hothorn2015ctree} which splits based on significance tests instead of information gain. 
These tree-based regression methods can fit very complex, sophisticated models. }

\subsubsection{Linear Model}
{The linear model is the simplest regression model.
Given a set of features $X_1, .. , X_n$, the linear model fits a function of the form $f(X_1, .. , X_n) = \beta_0 + \sum_{i = 1}^{n} \beta_i*X_i$, where the coefficients $\beta_i$ are real numbers. 
This project uses the default ordinary least squares (OLS) linear model from the SuperLearner package. 
Here, the optimal $\beta$ parameter vector is picked such that it minimizes the average squared error.
}

\subsection{SuperLearner}\label{sec:SL}
\textcolor{black}{
SuperLearner \citep{van2007super} is an ensemble learning algorithm that utilizes 10fCV.
Ensemble learning is an ML technique that combines multiple individual algorithms into a single one.
There are multiple methods to do this \citep{opitz1999popular}.
The advantage of such a method is that the shortcomings of one constituent algorithm can be mitigated by leveraging other algorithms. 
This overall has a positive impact, thereby boosting performance. 
In the case of SuperLearner, this ensemble is created by assigning coefficients to its constituent models. 
The coefficients are determined based on the RMSE of each algorithm in 10fCV.
Algorithms with lower RMSE are assigned higher coefficients.
The sum of all the coefficients always equals one, and they are greater than or equal to 0.
This ensemble created by SuperLearner intrinsically has the lowest RMSE among all its constituent models \citep{polley2010super}.
}


\subsection{SLOPE}\label{sec:SLOPE}
{Sorted L-One Penalized Estimation, or SLOPE \citep{bogdan2015slope}, is a supervised learning method for both prediction and feature selection. 
Given a vector $b$ $\in$ $\mathbb{R}^p$, and a sequence of non-negative scalars $\lambda_1 \ge \lambda_2 \ge \lambda_3 \ge ... \ge \lambda_p \ge 0$ where $\lambda \ne 0$, the Sorted L1-Norm is defined as $\lambda_1 b_1 + \lambda_2 b_2 + ... + \lambda_p b_p$. 
SLOPE is a natural extension of the LASSO method. 
It aims to minimize a cost function with the constraint that the Sorted L-1 Norm is less than or equal to a positive constant \citep{opitz1999popular}.
It does this by assigning larger weights for larger parameter values. 
We test this independent method that is not included in the SuperLearner ensemble. 
}


\subsection{Feature Engineering}\label{sec:featureEngineering}
{It is common in machine learning to create new predictors from the pre-existing features in the data sample \citep{ng2013machine}. 
\textcolor{black}{These techniques, called feature engineering, can help boost the performance of the final model.} 
The importance of feature engineering is clear when considering that the response variable may depend on combinations of the initial features that certain machine learning algorithms cannot learn. 
These include, but are not limited to, features such as ratios for neural networks, and count variables for tree algorithms \citep{Heaton_2016}. 
Continuing the work from \citet{dainotti2021predicting} where we considered models trained with the features present in 4LAC, here we extended the set of predictors.
\textcolor{black}{
Our engineered features are simple cross-multiplications of the existing features, as described in Sec. \ref{sec:sample}
to distinguish the feature sets, we denote the original features obtained from the 4LAC catalog as Order 1 (O1).
The cross multiplied features are called Order 2 (O2). 
Additionally, the feature selection with LASSO and the algorithms used \textcolor{black}{is done in the same way} for O1 and O2 cases.
The results are presented for both.}
}

\subsection{Bias Correction}\label{sec:BiasCorrection}
\textcolor{black}{
We define bias as the difference between the predicted and the observed values of a response variable.
Training models on an unbalanced sample often leads to predictions that have a high bias, although it should be noted that an unbalanced sample is not the sole reason for predictions to be biased.
This outcome can occur for other reasons such as the ML methods used, and the transformations applied to the predictors and the response variable.
In the case of an unbalanced sample, the model is better trained on the redshift range that contains more observations.
As a result, the model predicts more frequently in those particular redshift ranges and may provide erroneous results. 
}
\textcolor{black}{
\newline\indent
\textbf{We apply the technique of Optimal Transport for our bias correction. 
In this technique, the predicted and observed 1/(z+1) values are sorted in ascending order, and then a linear fit is obtained between them. 
The bias is corrected using the slope and intercept of this linear fit. 
The analysis is performed for results which are obtained in the case of O1 predictors with SuperLearner and is applied for BLLs and FSRQs independently. 
We correct the respective predictions using the following formula:
$$
C_{preds} = B + A * U_{phot}
$$
where $C_{preds}$ are the corrected 1/(z+1) predictions, $U_{phot}$ is the Superlearner 1/(z+1) predictions, $A$ and $B$ are the slope and intercept of the linear fits, respectively.
$C_{pred}$ values are then converted to the z-scale to obtain the bias corrected redshift estimates.
}
\newline\indent
The effect of these corrections is presented in Sec. \ref{sec:results_bias_correction}.
Following this procedure, we also correct the predictions on the generalization set and we present these results in Table \ref{tab:table3}, along with the predictions without the bias correction.
}

\section{Comparison}

{
{As mentioned, this study \textcolor{black}{aims at extending} the method developed by 
\citet{dainotti2021predicting}. Here we provide a comparison between the two studies.
\newline\indent
Previously, in \cite{dainotti2021predicting}, some of us worked with $2863$ sources from the 4LAC, comprising of $658$ FSRQs, $1067$ BLLs, $1074$ BCUs, and $64$ sources categorized as radio galaxies, Narrow line Seyferts, or other non-blazar AGNs. 
Out of those $2863$ sources, only BLLs and FSRQs, with no missing data, were used, resulting in $657$ AGNs for the training set and $73$ AGNs for the validation set. 
In the current study, the 4LAC catalog has been significantly expanded from $2863$ total AGNs to $3511$, an increment of $52\%$. 
This increment has been achieved by including the lower galactic latitude $\gamma$-ray loud AGNs.
Out of the total number of AGNs under consideration, $50.3\%$ have a measured redshift.
Thus, the first difference between the current and \cite{dainotti2021predicting} is the enlarged sample.
Indeed, in Sec. \ref{sec:sets}, we describe how we obtained a training set of $1112$ AGNs, a $52\%$ increase compared to \cite{dainotti2021predicting}. 
\newline \indent 
In addition, part of our extension of the work in \cite{dainotti2021predicting} is the consideration of new ML algorithms.
They had obtained the highest weights for the algorithms Extreme Gradient Boosting (XGBoost), Random Forest, Big LASSO, and Bayes Generalized Linear Model (Bayes GLM) using SuperLearner.
For this study, we obtain the highest weights for KSVM, Random Forest, Linear Model, Ranger, Cforest, and EARTH (see Sec. \ref{sec:ML}) within the same SuperLearner framework.
\newline \indent 
Previously, \cite{dainotti2021predicting} used the property $Highest\_Energy$ as one of the predictors and did not take into account for $Log\nu f\nu$, as this was already considered a cross-product term.
In this study, we are including $Log\nu f\nu$ and excluding $Highest\_Energy$ (see Sec. \ref{sec:sample} for additional details), along with introducing the use of O2 predictors (see Sec. \ref{sec:featureEngineering}). 
\newline\indent
In \cite{dainotti2021predicting}, LASSO feature selection was performed on the entire training set.
Here, we perform LASSO feature selection on a fold-by-fold basis inside the 10fCV leading to more accurate results (see Sec. \ref{sec:FeatureSelection}).
This procedure was applied in all the trails that use SuperLearner, regardless of the predictor set.
}}

\section{\textbf{Results}} \label{sec:results}
{In this study, we \textcolor{black}{tested multiple models with different parameter spaces} to obtain the best possible model for predicting the redshift of $\gamma$-ray loud AGNs.
For each of \textcolor{black}{the methods}, we perform 10fCV 100 times.
\textcolor{black}{The method which leads to the most promising results is the SuperLearner ensemble using O1 predictors}.
We then contrast this with \textbf{two} other experiments: SuperLearner ensemble with O2 predictors, 
and the SLOPE method with O1 predictors.
\newline\indent
The experiment with SuperLearner using O2 predictors is intended to test how feature engineering affects the predictive capabilities of the model. For this we have 78 O2 predictors compared to 12 O1 predictors.
\newline\indent
With SLOPE, we want to test how this single algorithm performs with respect to the SuperLearner ensemble. We are using the O1 predictor set with SLOPE and the same feature selection method using LASSO.
}

\subsection{SuperLearner ensemble with O1 predictors}\label{sec:O1_wo_MICE}
{Using the O1 predictor set with the SuperLearner ensemble presents us the best results amongst the experiments we tested.
}
\begin{figure}[H]
    \centering
    \includegraphics[width=0.49\textwidth]{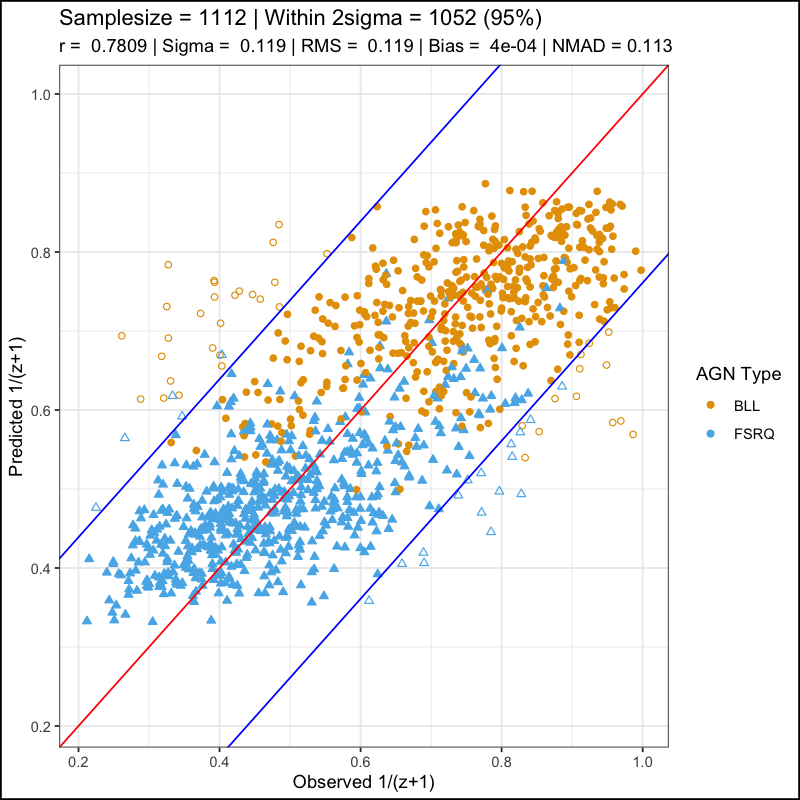} 
    \includegraphics[width=0.49\textwidth]{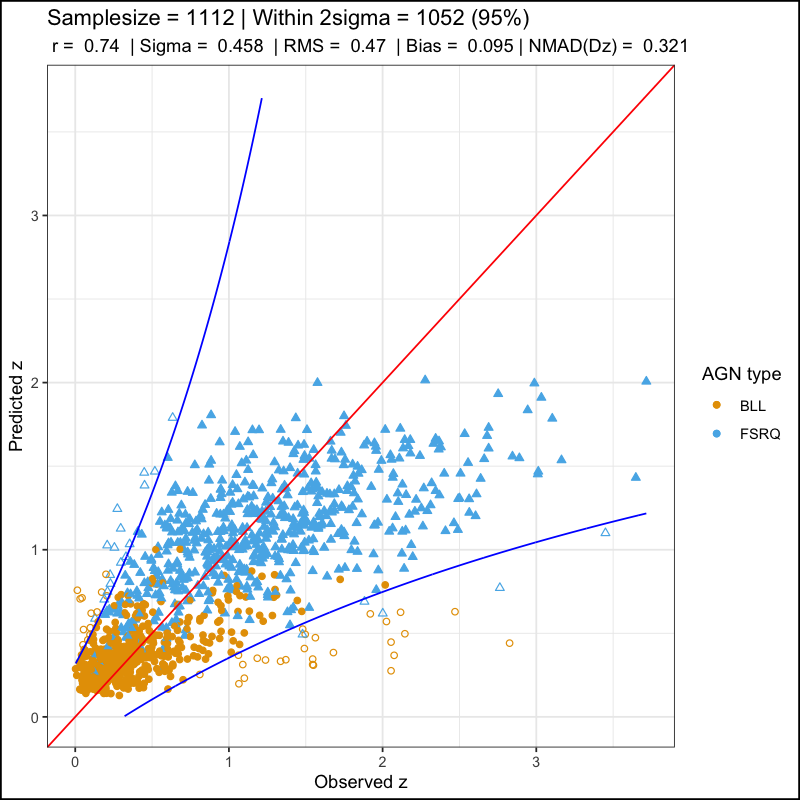}
    \includegraphics[width=0.49\textwidth]{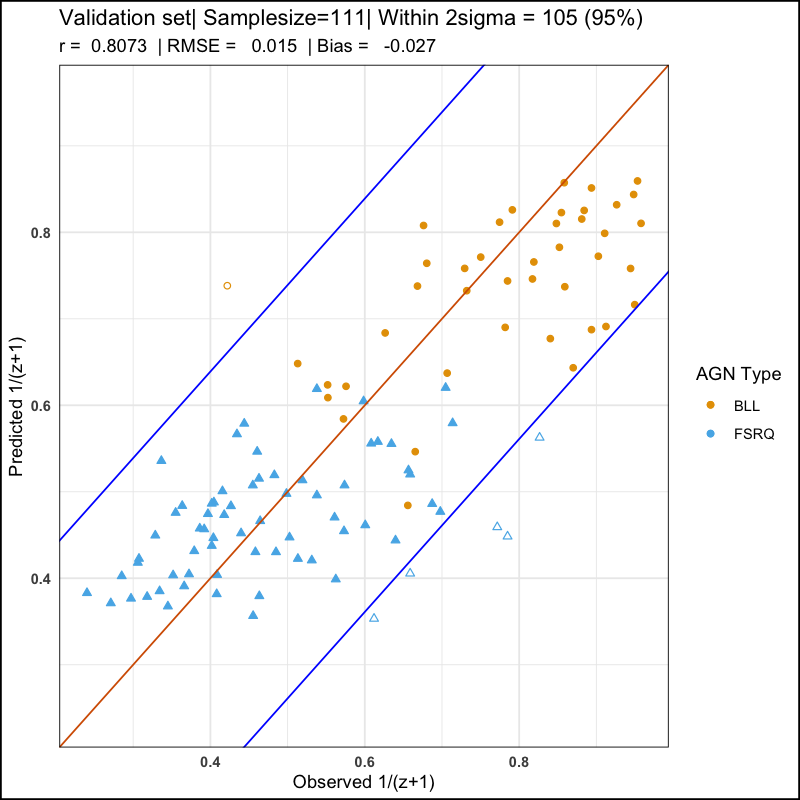}
    \includegraphics[width=0.49\textwidth]{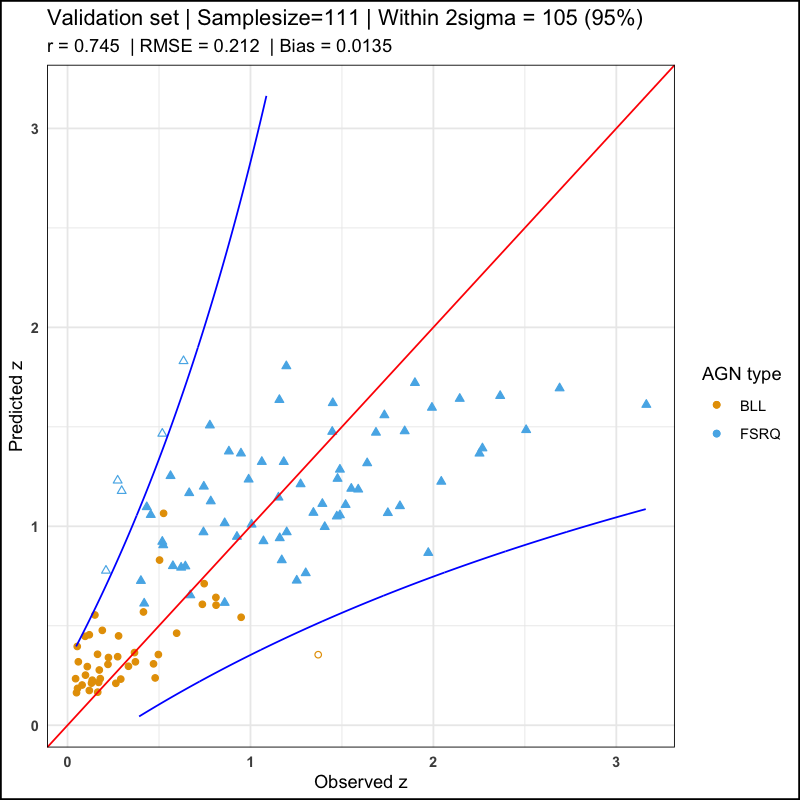}
    \caption{10fCV results.
    The orange filled circles represent the BLLs and the blue filled triangles represent the FSRQs. The empty circles and triangles represent the BLLs and FSRQs that are catastrophic outliers, respectively.
    Top left panel: cross-validation correlation plot in the 1/(z+1) scale.
    Top right panel: cross-validation correlation plot in the linear scale.
    Bottom left panel: Validation set correlation plot in the 1/(z+1) scale.
    Bottom right panel: Validation set correlation plot in the linear scale.
    }
    \label{fig:O1_wo_MICE_2}
\end{figure}

{The results obtained are shown in the Fig. \ref{fig:O1_wo_MICE_2}.
The top left and right panels show the cross-validated results, in the 1/(z+1) and linear scales, respectively.
In the 1/(z+1) scale we obtain a Pearson correlation of $78.1\%$ between the predicted and observed values, an RMSE of $0.119$, $\sigma$ of $0.119$, and $\sigma_{NMAD}$ of $0.113$. }
{In the linear scale, the correlation between predicted and observed redshift is $74\%$, RMSE is $0.467$, bias is $0.094$, and $\sigma_{NMAD}$ is $0.321$. 
We obtain a low catastrophic outlier percentage of $5\%$: $1052$ AGNs are predicted within the $2\sigma$ range. 
}

\textcolor{black}{
The lines in blue depict the 2$\sigma$ curves for each plot, where the $\sigma$ is calculated in the 1/(z+1) scale as follows:
$$ \frac{1}{z_p + 1} = \frac{1}{z_s + 1} \pm 2\sigma, $$
\\
where $z_s$ is the spectroscopic redshift and $z_p$ is the photometric redshift.
{The choice of our scaling causes the 2$\sigma$ curve to not be a straight line on the linear scale. 
Thus, we use the formula mentioned below to draw them in the linear-scale.
$$
z_p = z_s \left[ \frac{1 \pm 2\sigma(z_p + 1)}{1 \mp 2\sigma} \right] \pm \frac{2\sigma}{1 \mp 2\sigma.} 
$$}
}
\begin{figure}[H]
    \centering
    \includegraphics[width=0.49\textwidth]{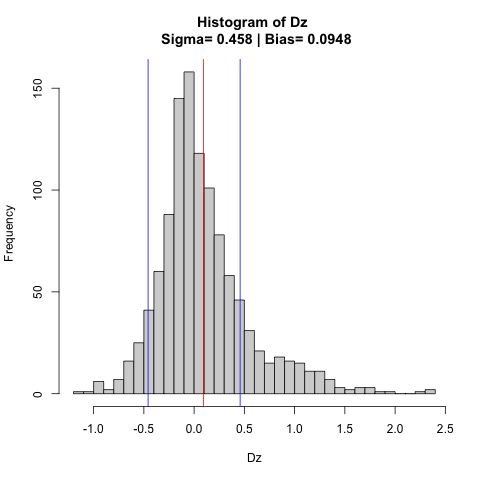} 
    \includegraphics[width=0.49\textwidth]{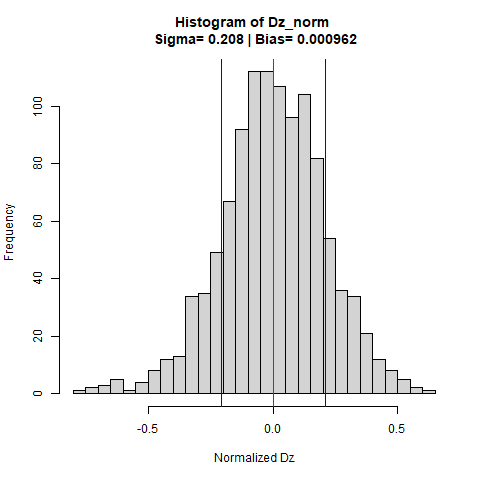}
    \includegraphics[width = 0.49\textwidth]{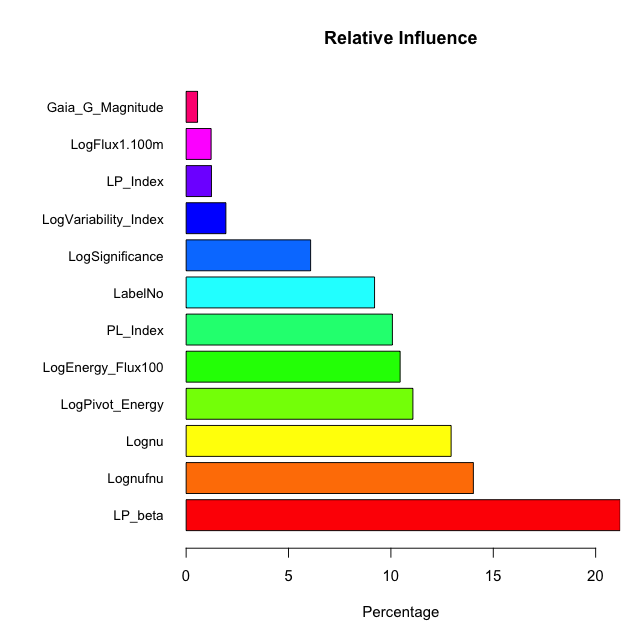}
    \includegraphics[width=0.49\textwidth]{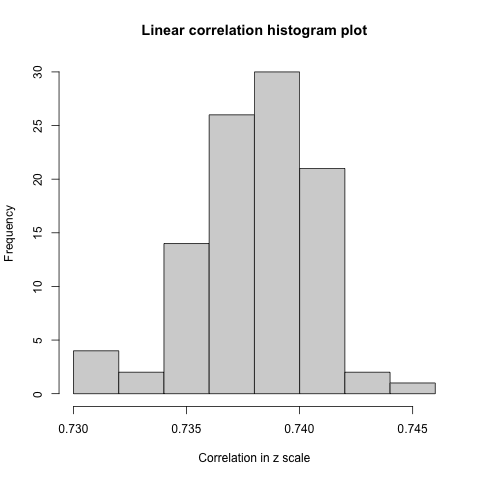}
    \caption{
    Top Left Panel: Distribution of $\Delta z$ with average bias (red) and sigma lines (blue).
    Top Right Panel: Distribution of the $\Delta z_{norm}$ with the average bias (red) and sigma values (blue).
    Bottom Left Panel: Relative influence of the O1 predictors.
    Bottom right panel: Distribution of the correlations in the linear scale from the 100 iterations of 10fCV.
    }
    \label{fig:O1_wo_MICE_spreads}
\end{figure}

{In the Fig. \ref{fig:O1_wo_MICE_spreads} the top left and right panels show the histogram distribution of the $\Delta$z and $\Delta$z$_{norm}$. 
The bias is denoted by the red line and the blue lines show the $\sigma$ values. 
The bottom left panel shows the relative influence of our predictors with $LP\_\beta$ with the highest relative influence, followed by the $Log\nu$ and $Log\nu f\nu$. 
The categorical predictor $LabelNo$, which distinguishes between BLL and FSRQ, is the 7th most influential predictor.
These relative influence values presented here are an average over 100 iterations. 
Such a step is necessary as it serves to derandomize and stabilize the relative influence results, and thus they can be compared with \cite{dainotti2021predicting}.
Finally, the bottom right panel shows the distribution of the correlation value between observed and predicted redshift across the 100 iterations of SuperLearner. 
As shown, these correlations lie completely within 0.73 to 0.745, with the majority falling between 0.735 and 0.74.
}


\begin{figure}[H]
    \centering

    \includegraphics[width=0.49\textwidth]{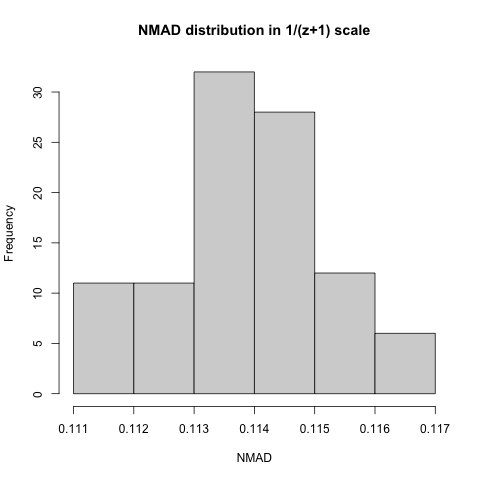} 
    \includegraphics[width=0.49\textwidth]{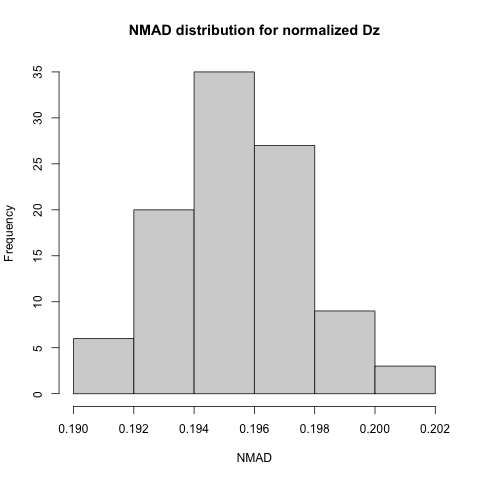}
    \includegraphics[width=0.49\textwidth]{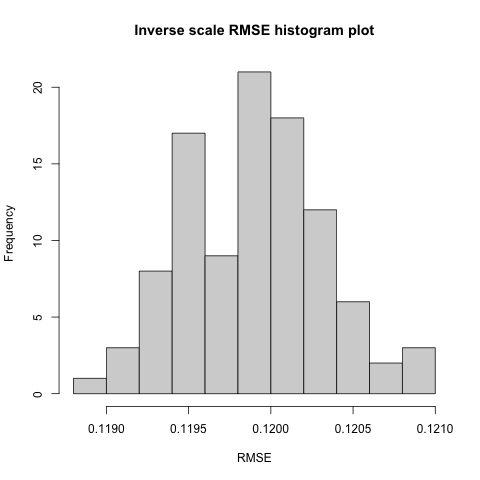} 
    \includegraphics[width=0.49\textwidth]{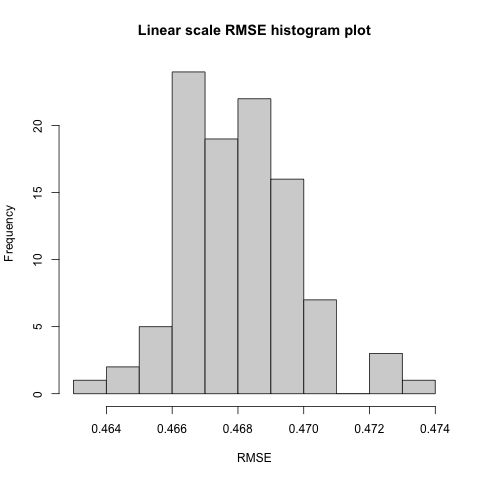}
    \caption{These distribution plots are derived from the 100 iterations of 10fCV. They represent how dispersed these metrics are for the data and algorithms at hand.
    Top left panel: Distribution of $\sigma_{NMAD}$ in the 1/(z+1) scale. The $\sigma_{NMAD}$ is written as NMAD on the plots.
    Top right panel: Distribution of the $\sigma_{NMAD}$ in the linear scale.
    Bottom left panel: Distribution of the RMSE in the 1/(z+1) scale.
    Bottom right panel: Distribution of the RMSE in the linear scale.}
    \label{fig:O1_wo_MICE_3}
\end{figure}

\textcolor{black}{
In Fig. \ref{fig:O1_wo_MICE_3} we are presenting the distributions of two parameters, namely RMSE and $\sigma_{NMAD}$.
The top left and right panels show the distribution of $\sigma_{NMAD}$ in the 1/(z+1) and the linear scales, respectively. 
Similarly, the bottom left and right panels show the distribution of RMSE in 1/(z+1) and the linear scales, respectively. 
}

\subsection{Comparison with the other methods}
In this section, we present a comparative analysis between the O1 predictors with SuperLearner and the \textbf{two} experiments we mentioned previously.
\newline\indent
\newline\indent
When using the O2 predictor set with the SuperLearner ensemble we obtain comparable results to the O1 case.
In the 1/(z+1) scale, the correlation is 0.782, which is within 1\%. Similarly, we obtain a RMSE of 0.119, bias of 4.5$\times10^{-4}$, and $\sigma_{NMAD}$ of 0.114. 
\newline\indent
In the linear scale, the correlation is 0.74, which is equal to the value obtained with the O1 predictors. The RMSE is 0.463, which is a 1\% improvement. The bias obtained with O2 is 0.088, which is similarly a 1\% improvement. The $\sigma_{NMAD}$ obtained is 0.328.
\newline\indent
\newline\indent
{As mentioned previously, we also used the SLOPE method. Compared to the SuperLearner ensemble with O1 predictors, it performs nearly as well despite being a singular algorithm. 
Specifically, we obtain a Pearson correlation of 76\% in the 1/(z+1) scale, a RMSE of 0.124, bias of 3$\times10^{-5}$, and $\sigma_{NMAD}$ of 0.12. 
\newline\indent
In the linear scale, the correlation is 72\%, the RMSE is 0.479, bias is 0.1 and $\sigma_{NMAD}$ is 0.348. 
Compared to the O1 case, in the linear scale, the R is down by 2.7\%, RMSE is up by 2.5\%, and the $\sigma_{NMAD}$ is up by 8.4\%.
In both the linear scale and the 1/(z+1) scale the O1 predictors with SuperLearner outperforms SLOPE, although by only a small margin.
}
\newline\indent
\newline\indent
All the results are presented in Table. \ref{tab:comparision1} and Table. \ref{tab:comparision2}, for the 1/(z+1) and the linear scales, respectively.

\begin{table}[]
    \centering
    \begin{tabular}{c|c|c|c}
        Metric & SL with O1 & SL with O2 & SLOPE with O1 \\
        \hline
        \hline
        R & 0.781 & 0.782 & 0.762 \\
        \hline
        RMSE & 0.119 & 0.119 & 0.124  \\
        \hline
        Bias & $4\times10^{-4}$ & $4.5\times10^{-4}$ & $3\times10^{-5}$ \\
        \hline
        $\sigma_{NMAD}$ & 0.113 & 0.114 & 0.12 \\
        \hline
        $\sigma$ & 0.119 & 0.119 & 0.124 \\
    \end{tabular}
    \caption{Metrics to compare the different experiments performed in the 1/(z+1) scale.}
    \label{tab:comparision1}
\end{table}

\begin{table}[]
    \centering
    \begin{tabular}{c|c|c|c}
        Metric & SL with O1 & SL with O2 & SLOPE with O1 \\
        \hline
        \hline
        R & 0.74 & 0.74 & 0.72 \\
        \hline
        RMSE ($\Delta z$) & 0.467 & 0.463 & 0.479 \\
        \hline
        Bias ($\Delta z$) & 0.095 & 0.089 & 0.1 \\
        \hline
        $\sigma_{NMAD}$ ($\Delta z$) & 0.321 & 0.328 & 0.348\\
        \hline
        $\sigma$ ($\Delta z$) & 0.458 & 0.455 & 0.47\\
        \hline
        Bias ($\Delta z_{norm}$) & $9.6\times10^{-4}$ & $-5.1\times10^{-4}$ & $6.9\times10^{-4}$ \\
        \hline
        $\sigma_{NMAD}$ ($\Delta z_{norm}$) & 0.195 & 0.195 & 0.205 \\
        \hline
        $\sigma$ ($\Delta z_{norm}$) & 0.208 & 0.209 & 0.217 
    \end{tabular}
    \caption{Metrics to compare the different experiments performed in the linear scale.}
    \label{tab:comparision2}
\end{table}




\subsection{Prediction on the generalization set}
\begin{figure}
    \centering
	\includegraphics[width=0.8\textwidth]{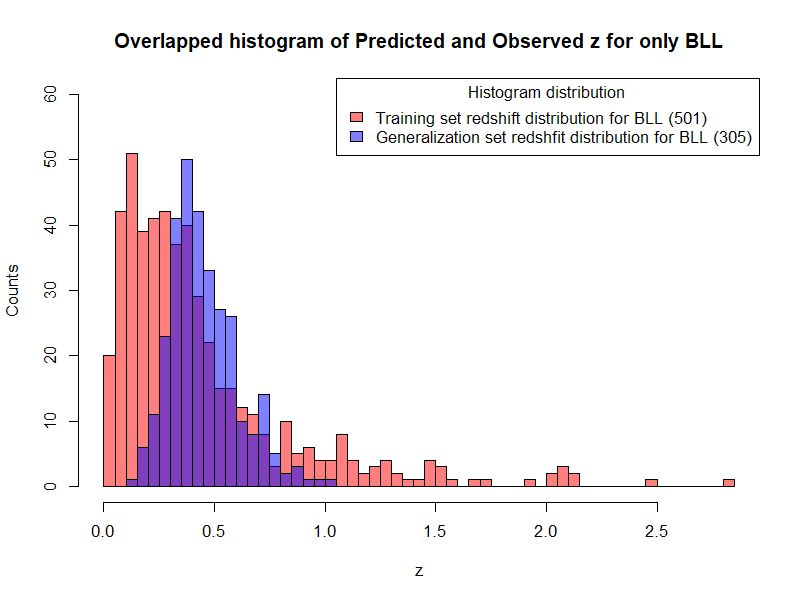}
    \caption{The distribution of the predicted redshifts for the generalization set from the model trained on O1 predictors, overlapped with the redshift distribution of the BLLs in the training set.}
    \label{fig:genset_preds}
\end{figure}

\textcolor{black}{
As described in Sec. \ref{sec:sets} the generalization set consists of AGNs that do not have a measured redshift. 
\begin{figure}
    \centering
	\includegraphics[width=\textwidth]{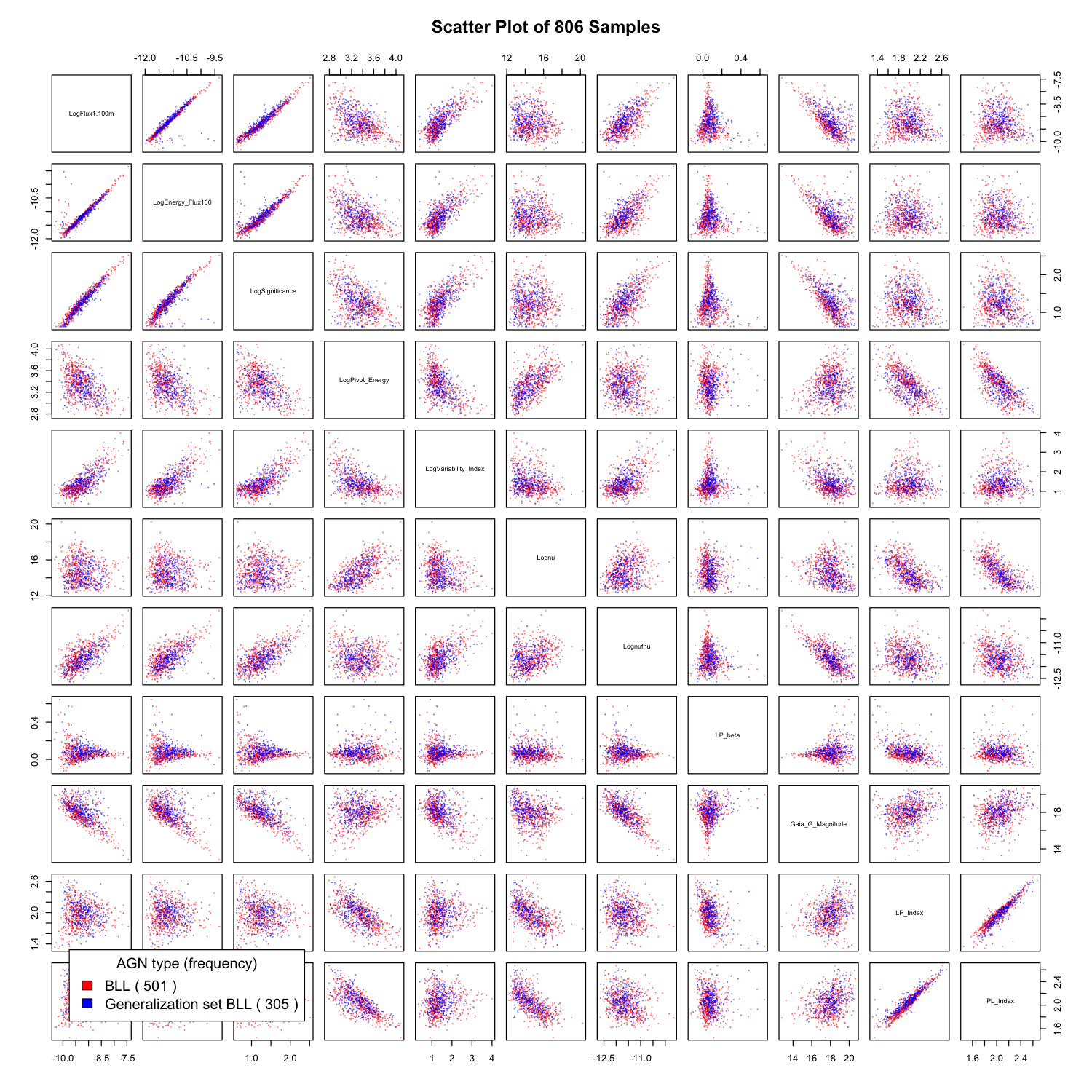}
    \caption{The scatter matrix plot for the BLLs in the training (red) and generalization (blue) samples.}
    \label{fig:BLL_Scatter_Matrix}
\end{figure}
When predicting on this set it is crucial to ensure that the data points fall within the parameter space of the trained model.
Otherwise, the model will extrapolate those predictions and our confidence in them will be drastically reduced.
We ensure this does not happen by only keeping those AGNs in the generalization set whose maximum and minimum values for each predictor fall within the range of the same predictor in the training set.
\textbf{This is demonstrated in Fig. \ref{fig:BLL_Scatter_Matrix}, where the blue points (Generaliztion set BLLs) lie within the range of the red points (Training set BLLs).}
This leaves us with 305 AGNs, all of which are BLLs.
\newline\indent
In Fig. \ref{fig:genset_preds}, we show the distribution of the predicted values of 305 BLLs, overlapped with the redshift distribution of the BLLs in the training sample.
These predictions are made by using the O1 predictor set with the SuperLearner ensemble.
}

\subsubsection{Comparative results}

\begin{figure}[H]
    \centering
	\includegraphics[width=0.49\textwidth]{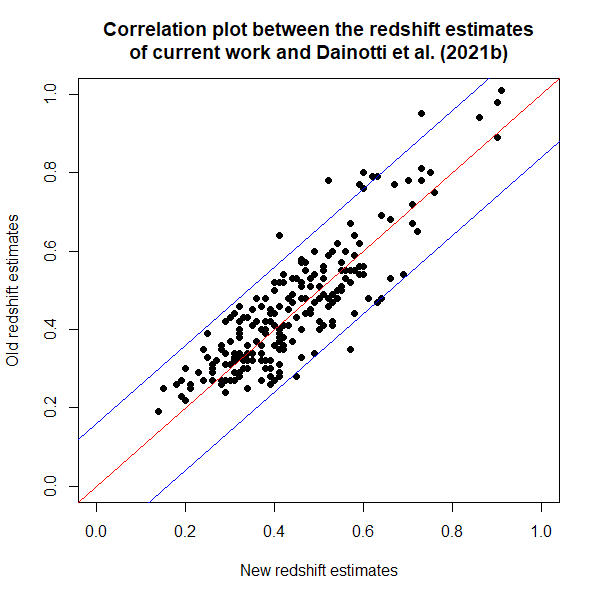}
    \caption{Comparision between the redshift estimates of 240 generalization set AGNs, common between this study and \cite{dainotti2021predicting}.
    We obtain a correlation of 0.87, RMSE of 0.078 and $\sigma$ of 0.079. 
    }
    \label{fig:genset_comparison}
\end{figure}

Here we consider the intersection of the generalization sets of both the current study and \cite{dainotti2021predicting}, which consists of 240 AGNs.
By plotting the estimated redshifts of these AGNs from the two studies, we see that there is a Pearson correlation of 87\%.
This correlation plot is presented in Fig. \ref{fig:genset_comparison}. 
We also observe a low RMSE and $\sigma$ of 0.078 and 0.079, respectively. 
The red line denotes the equality line, while the blue lines denote the 2$\sigma$ error curves. 
Based on the position of the points in the plot, we note here that the redshift estimate of only 8 AGNs (3\%) lie beyond the 2$\sigma$ error limit. 
It should be noted that both these sets of estimates have not been bias corrected.

\subsection{Bias Correction for the Generalization set}\label{sec:results_bias_correction}

\begin{figure}[H]
    \centering
    \includegraphics[width=0.49\textwidth]{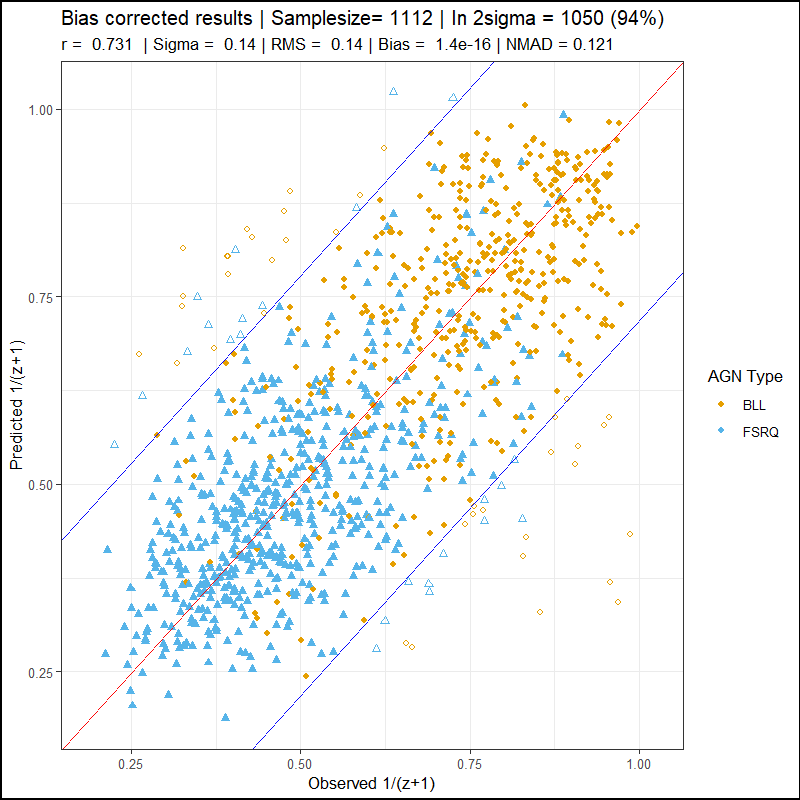}
    \includegraphics[width=0.49\textwidth]{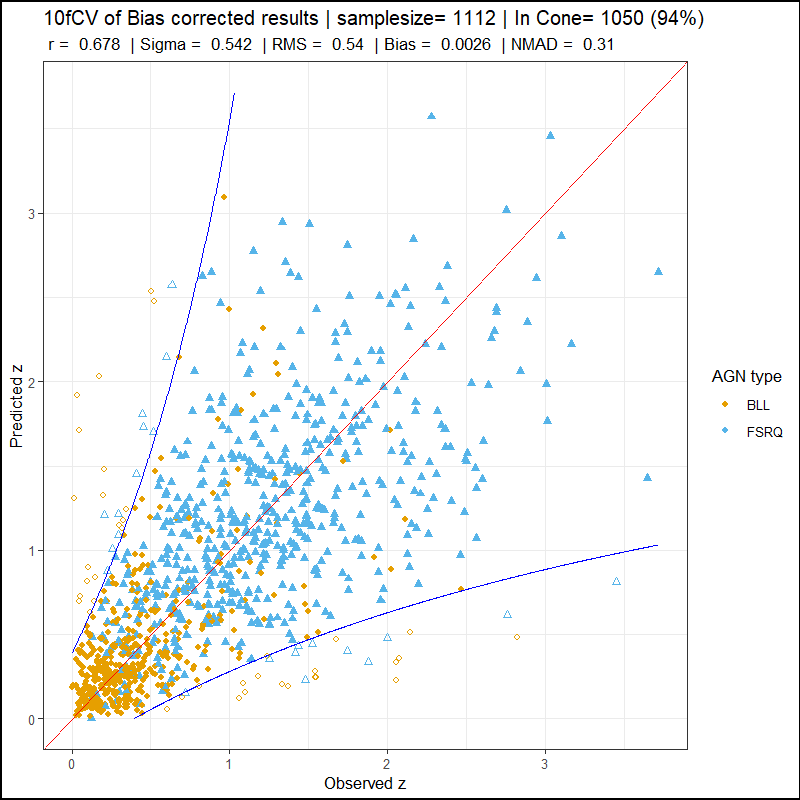}
    
	\caption{The bias-corrected predictions of the 10fCV.
    The symbols here retain the same meaning as those in Fig. \ref{fig:O1_wo_MICE_2}.
    Left Panel: Correlation plot between the bias-corrected prediction and observed redshift in the 1/(z+1) scale. 
    Right Panel: Correlation plot between the bias-corrected prediction and observed redshift in the linear scale}
    \label{fig:CV_BC}
\end{figure}

\begin{figure}[H]
    \centering
	\includegraphics[width=0.8\textwidth]{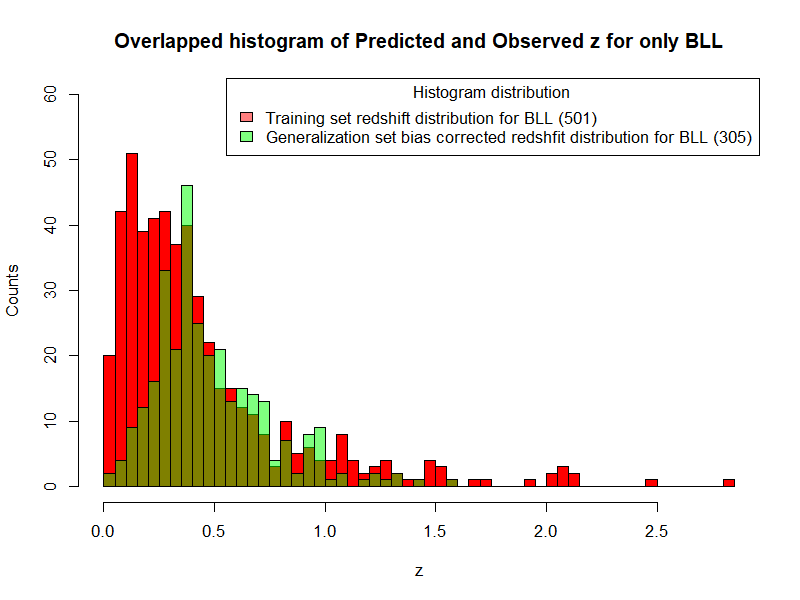}
    \caption{Histogram distributions of bias-corrected predictions of the generalization set AGNs, overlapped with the training set AGNs.}
    \label{fig:BC_preds}
\end{figure}

\textcolor{black}{
Following the bias-correction described in Sec. \ref{sec:BiasCorrection}, the modified 10fCV results are presented in Fig. \ref{fig:CV_BC}.
The left and right panels of Fig. \ref{fig:OT} present the optimal transport linear fit of BLLs and FSRQs, respectively.
\begin{figure}[H]
    \centering
    \includegraphics[width=0.49\textwidth]{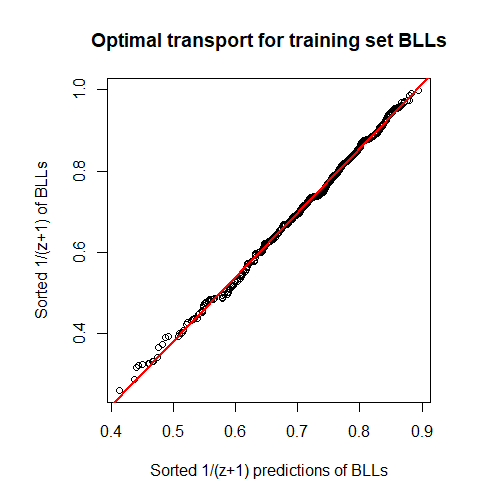}
    \includegraphics[width=0.49\textwidth]{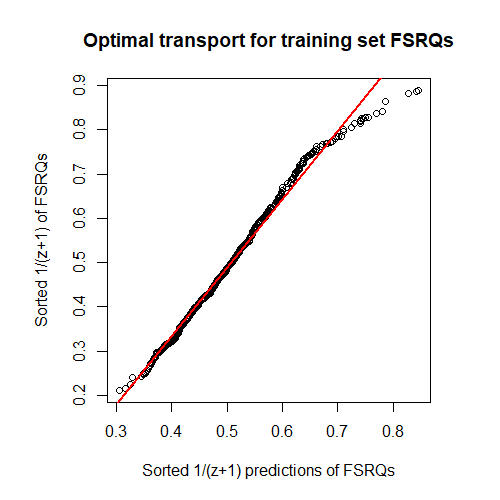}
	\caption{Optimal transport linear fits of BLLs, left panel, and FSRQs, right panel.}
    \label{fig:OT}
\end{figure}
We note that there is a \textbf{8.3}\% reduction in the correlation in the linear scale, \textbf{15.6}\% increase in the RMSE, and a \textbf{3.4\% decrease} in the $\sigma_{NMAD}$.
\textbf{We also see an increase in catastropic outliers from 5\% to 6\%. However, in actual numbers there are only two additional AGNs that are outside the 2$\sigma$ cones, as compared to the O1 scenario.}
These results are quite expected, since with bias correction we are reducing the accuracy in the bulk of the redshift distribution, in order to get better results for the tails of the distribution.
The application of this bias correction to the predictions on the generalization set lead to the results shown in Fig. \ref{fig:BC_preds}.
We present the full set of bias-corrected predictions in Table. \ref{tab:table3}.
}

{
}

\section{\textbf{Conclusions \& Discussion}} \label{sec:conclusion}
{In \cite{dainotti2021predicting}, some of us had obtained a Pearson correlation of 0.71 and an RMSE of 0.434 in the linear scale. In this study, using the O1 predictors with the SuperLearner we achieve a Pearson correlation of 0.74 and an RMSE of 0.467. Specifically, an improvement of 4.2\% in the correlation and an increase of 7.6\% in RMSE.
At the same time, the percentage of catastrophic outliers remains the same at 5\% between this study and  \cite{dainotti2021predicting}.
}
\newline\indent
\textcolor{black}{
We found that using O2 predictors leads to performance equivalent to O1 predictors.
{Thus, we can deduce that O1 predictors contain all the information needed to predict the redshift.
This is evident in Fig. \ref{fig:O2_relinf} as the top ten O2 predictors with the highest relative influence are either O1 predictors or cross products of other high influence O1 predictors (see bottom left panel of Fig. \ref{fig:O1_wo_MICE_spreads}).
Such an outcome could possibly be exclusive to the 4LAC properties, and indeed cross-product terms in other data sets might lead to better performance.}
}
\newline\indent
\textcolor{black}{
Additionally we tested SLOPE independently of SuperLearner.
On average we observe that the results from SLOPE are comparable to those from the ensemble via the metrics presented in Table \ref{tab:comparision1}.
However, the advantage of SLOPE is that it works much faster than the SuperLearner ensemble, along with the added benefit of high interpretability of the algorithm.
Hence, when the requirement is to obtain estimates quickly, SLOPE can be used as a reliable alternative.
}
\newline
\newline\indent
Here we summarise about the predictors used in our study.
\begin{enumerate}
    \item $LP\_\beta$, $Log\nu f\nu$, $Log\nu$, $LogPivot\_Energy$, $LogEnergy\_Flux$, $PL\_Index$, $LabelNo$, and $LogSignificance$ have highest influence in this study.
    
    \item $LP\_\beta$, $LogPivot\_Energy$, $LogSignificance$, $LogEnergy\_Flux$, $Gaia\_G\_Magnitude$, $Log\nu$, and $Log\_Highest\_Energy$ had the highest influence in \cite{dainotti2021predicting}.
    
    \item In this study and \cite{dainotti2021predicting}, $LP\_\beta$ has the highest influence at 25\%.
    
    \item With O2 predictors, $LP\_\beta$ has a relative influence of 3\% at the 10th position.
    Three O2 predictors more influential than $LP\_\beta$ are cross-products of it. These are $LP\_\beta^2$, $LogSignificance*LP\_\beta$ and $LogVariability\_Index*LP\_\beta$, with $LogVariability\_Index*LP\_\beta$ being the second most influential at 10\%.
    
    \item Hence, we can conclude that for redshifts estimation of 4LAC AGNs, $LP\_\beta$ is crucial.
    
    \item $Gaia\_G\_Magnitude$ has a relative influence of 1\% at 12th position.
    In \cite{dainotti2021predicting} however it was the 5th most influential at 12\% relative influence. This shift in the relative influence of $Gaia\_G\_Magnitude$ can be attributed to the enlarged data set. However, this is not conclusive and merits further investigation. 
    
    \item $PL\_Index$ has the highest influence when paired with O2 predictors at 15\%. With O1 predictors, it was the 6th most influential at 11\% influence.
    

\end{enumerate}

\textcolor{black}{
Finally, to make the 4LAC more complete, we present two lists containing the predictions on the generalization set.
One contains the predictions from the SuperLearner ensemble with O1 predictors (shown in Fig. \ref{fig:genset_preds}) and the second contains the bias-corrected results (shown in Fig. \ref{fig:BC_preds}).
{Additionally, we provide a comparison between the redshift estimates of 240 AGNs that are common between the generalization sets of this study and \cite{dainotti2021predicting} (see Fig. \ref{fig:genset_comparison}). 
These two results have a Pearson correlation of 0.87, RMSE and $\sigma$ of 0.078 and 0.079, respectively.
}}
\textcolor{black}{
\newline\indent
This is the most complete investigation into the predictive capabilities of the 4LAC dealing with redshift estimation. 
Compared to the \cite{dainotti2021predicting}, we have been able to maintain the predictive capabilities of our methodology with an even larger data sample. These estimates will likely serve as a reference point for the community to be used as tools for cosmological investigations. The final aim is to use the estimation of the redshift to investigate AGN with interesting properties.
}
\newline\indent

\section{Appendix}\label{sec:appendix}

Here we are presenting Table. \ref{tab:table3} containing the predictions of fifty generalization set AGNs. 
It contains the names of AGNs, as assigned in the 4LAC catalog, followed by the predicted redshift from the SuperLearner ensemble with O1, and finally the bias-corrected redshifts. The predictions for the complete generalization set can be found in the data file provided.

\begin{table}[]
    \centering
    \begin{tabular}{|l|c|c|}
    \hline
        Name & Predicted $z_{pred}$ & Bias-corrected $z_{pred}$ \\
        \hline
  4FGL J0009.3+5030 &0.57&0.62
\\4FGL J0001.2-0747 &0.39&0.39
\\4FGL J0013.1-3955 &0.7&0.90
\\4FGL J0019.3-8152 &0.24&0.17
\\4FGL J0019.6+2022 &0.79&1.09
\\4FGL J0021.5-2552 &0.48&0.51
\\4FGL J0022.1-1854 &0.57&0.65
\\4FGL J0022.5+0608 &0.76&1.03
\\4FGL J0023.9+1603 &0.63&0.78
\\4FGL J0029.0-7044 &0.59&0.72
\\4FGL J0031.3+0726 &0.32&0.27
\\4FGL J0045.3+2128 &0.34&0.28
\\4FGL J0056.4-2118 &0.51&0.54
\\4FGL J0058.0-3233 &0.54&0.62
\\4FGL J0058.3+1723 &0.37&0.36
\\4FGL J0107.4+0334 &0.4&0.48
\\4FGL J0112.8-7506 &0.39&0.38
\\4FGL J0115.6+0356 &0.48&0.52
\\4FGL J0116.0-2745 &0.31&0.26
\\4FGL J0116.2-6153 &0.39&0.37
\\4FGL J0120.4-2701 &0.34&0.28
\\4FGL J0125.3-2548 &0.71&0.95
\\4FGL J0127.2+0324 &0.34&0.29
\\4FGL J0132.8-4413 &0.37&0.35
\\4FGL J0136.5+3906 &0.34&0.28
\\4FGL J0138.0+2247 &0.26&0.19
\\4FGL J0139.0+2601 &0.33&0.29
\\4FGL J0142.7-0543 &0.36&0.32
\\4FGL J0143.7-5846 &0.21&0.11
\\4FGL J0144.6+2705 &0.73&0.95
\\4FGL J0153.0+7517 &0.31&0.26
\\4FGL J0155.0+4433 &0.54&0.62
\\4FGL J0158.5-3932 &0.4&0.39
\\4FGL J0159.7-2740 &0.39&0.37
\\4FGL J0202.7+4204 &0.51&0.57
\\4FGL J0203.6+7233 &0.47&0.51
\\4FGL J0208.3-6838 &0.44&0.46
\\4FGL J0209.3+4449 &0.29&0.23
\\4FGL J0219.5+0724 &0.42&0.44
\\4FGL J0226.5-4441 &0.32&0.28
\\4FGL J0232.9+2608 &0.35&0.30
\\4FGL J0241.0-0505 &0.58&0.69
\\4FGL J0243.4+7119 &0.66&0.84
\\4FGL J0245.1-0257 &0.45&0.48
\\4FGL J0258.1+2030 &0.6&0.73
\\4FGL J0304.9-0606 &0.48&0.50
\\4FGL J0309.7-0745 &0.32&0.26
\\4FGL J0310.6-5017 &0.28&0.21
\\4FGL J0312.5-2221 &0.37&0.34
\\4FGL J0314.3-5103 &0.55&0.64
\\4FGL J0316.2-6437 &0.41&0.39
\\
    \hline
    \end{tabular}
    \caption{List of predictions for the generalization set. The first column has the names of the AGNs as they appear in the 4LAC catalog. Second column contains the redshift predictions from the SuperLearner ensemble using O1 predictors. Third column contains the same predictions with bias correction. Here $z_{pred}$ denotes the predicted redshift.}
    \label{tab:table3}
\end{table}

\begin{figure}[H]
    \centering
	\includegraphics[width=0.5\textwidth]{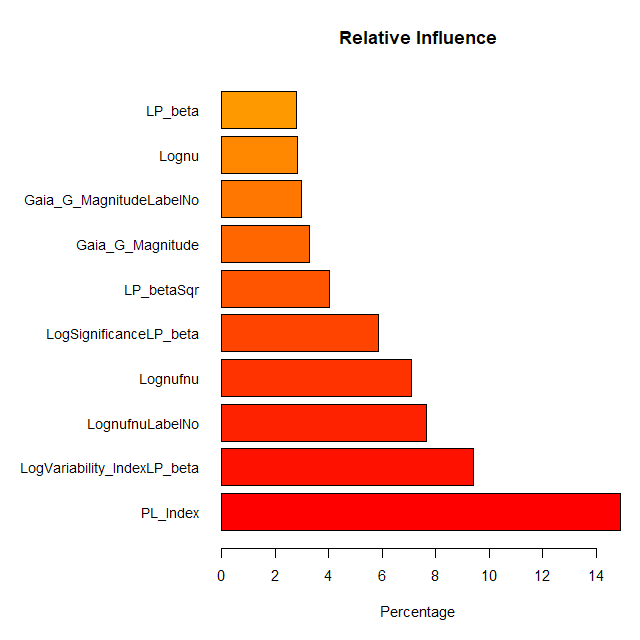}
    \caption{This plot presents the relative influence of top 10 the O2 predictors. }
    \label{fig:O2_relinf}
\end{figure}

The relative influence of the top 10 O2 predictors are shown in Fig. \ref{fig:O2_relinf}

\textbf{
Due to BLLs being the major category of AGNs present in the generalization set of the 4LAC catalog, we here present the distribution of predicted redshift made from a Superlearner model trained exclusively on the BLLs of the training set (501 BLL). 
We perform the Kolmogorov Smirnov (KS) test to compare the distribution of the predicted redshifts of generalization samples of BLLs, obtained from the model trained only on BLLs and the O1 SuperLearner model. 
We conclude based on the KS test (p-value=0.66) that we cannot reject the null hypothesis that the two distributions are drawn from the same population.
Thus, we have stable results, which are resistant to the construction of the training sample, either with only BLLs or with the total 4LAC sample. 
}
\begin{figure}[H]
    \centering
	\includegraphics[width=0.8\textwidth]{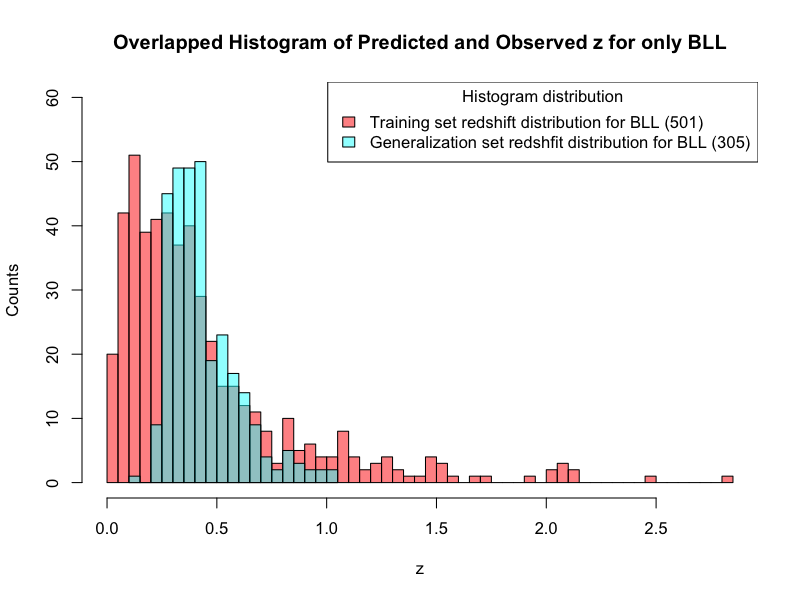}
    \caption{Distribution of the redshifts of the training data (red) and the predicted redshifts of the generalization set, from a Superlearner trained only on BLLs (turquoise).}
    \label{fig:hist_trained_on_BLLs}
\end{figure}

\acknowledgments
 This work presents results from the European Space Agency (ESA) space mission, Gaia. Gaia data are being processed by the Gaia Data Processing and Analysis Consortium (DPAC). Funding for the DPAC is provided by national institutions, in particular the institutions participating in the Gaia MultiLateral Agreement (MLA). The Gaia mission website is https://www.cosmos.esa.int/gaia. The Gaia archive website is https://archives.esac.esa.int/gaia.
M.G.D. thanks Trevor Hastie for the interesting discussion on overfitting problems.
We also thank Raymond Wayne for the initial computation and discussions about balanced sampling techniques which will be implemented in subsequent papers. 
This research was supported by the Polish National Science Centre
grant UMO-2018/30/M/ST9/00757 and by Polish Ministry of Science and
Higher Education grant DIR/WK/2018/12.
We would also like to acknowledge Enrico Rinaldi and Biagio De Simone for their invaluable discussions regarding bias correction and SLOPE implementation.

\bibliography{refs}

\end{document}